\documentclass[12pt]{article}

\usepackage[english]{babel}
\usepackage{booktabs}
\usepackage{array} 
\usepackage{authblk}
\usepackage[letterpaper,top=2cm,bottom=2cm,left=3cm,right=3cm,marginparwidth=1.75cm]{geometry}
\usepackage{setspace}
\usepackage{caption}

\usepackage[backend=biber,style=numeric-comp,sorting=none,maxnames=3,minnames=1]{biblatex}
\usepackage{amsmath}
\usepackage{graphicx}
\usepackage[colorlinks=true, allcolors=blue]{hyperref}
\usepackage{float}
\newcommand*\samethanks[1][\value{footnote}]{\footnotemark[#1]}

\title{Evaluating The Performance of Using Large Language Models to Automate Summarization of CT Simulation Orders in Radiation Oncology}
\author[1]{Meiyun Cao\thanks{Co-first author}}
\author[2]{Shaw Hu\samethanks} 
\author[1]{Jason Sharp}
\author[1]{Edward Clouser}
\author[1]{Jason Holmes}
\author[1]{Linda L. Lam}
\author[1]{Xiaoning Ding}
\author[1]{Diego Santos Toesca}
\author[1]{Wendy S. Lindholm}
\author[1]{Samir H. Patel}
\author[1]{Sujay A. Vora}
\author[1]{Peilong Wang\thanks{Co-corresponding author}} 
\author[1]{Wei Liu\samethanks}

\affil[1]{Department of Radiation Oncology, Mayo Clinic, Phoenix, AZ 85054}
\affil[2]{Material Science and Engineering, George Washington University, Washington, DC, 20052}

\date{}

\begin{document}
\singlespacing
\maketitle
\begin{abstract}
Purpose: In the current clinical workflow of radiation oncology departments, therapists manually summarize CT simulation orders into summaries before CT simulation is performed. This process increases the workload, introduces variability in documentation quality, and is prone to human errors. To address these challenges, this study aims to use a large language model (LLM) to automate the generation of summaries from the CT simulation orders and evaluate its performance. 

Materials and Methods: A total of 607 CT simulation orders for patients were collected from the Aria database at our institution.  A locally hosted Llama 3.1 405B model, accessed via the Application Programming Interface (API) service, was used to extract keywords from the CT simulation orders and generate summaries. The downloaded CT simulation orders were categorized into seven groups based on treatment modalities and disease sites. For each group, a customized instruction prompt was developed collaboratively with therapists to guide the Llama 3.1 405B model in generating summaries. The ground truth for the corresponding summaries was manually derived by carefully reviewing each CT simulation order and subsequently verified by therapists. The accuracy of the LLM-generated summaries was evaluated by therapists using the verified ground truth as a reference.  

Results: Over 98\% of the LLM-generated summaries aligned with the manually generated ground truth in terms of accuracy. Our evaluations showed an improved consistency in format and enhanced readability of the LLM-generated summaries compared to the corresponding therapists-generated summaries. This automated approach demonstrated a consistent performance across all groups, regardless of modality or disease site. 

Conclusions: This study demonstrated the high precision and consistency of the Llama 3.1 405B model in extracting keywords and summarizing CT simulation orders, suggesting that LLMs have great potential to help with this task, reduce the workload of therapists and improve workflow efficiency.
\end{abstract}

\section{Introduction}
In recent years, artificial intelligence (AI) has revolutionized a wide range of fields. Built on transformer architecture \cite{vaswani2023attentionneed} and trained on a vast corpora of text data, large language models (LLMs), such as ChatGPT (OpenAI, San Francisco, CA) \cite{chatgpt_webpage}, show the ability to analyze complex text information \cite{Liu2024a, holmes2024benchmarking, hao2024retrospective}, potentially assisting in decision-making \cite{prabhod2023integrating, vavekanand2024large, karttunen2023large, wang2024fine}. 
Meanwhile, commercially available LLMs have limitations, particularly in healthcare \cite{zhang2024potential, liuzhengliang2023, Wu2023ExploringTT}, where patient health information (PHI) protection is a major concern. The costs associated with large-scale API queries can also pose an additional financial charge to clinics.

In radiation oncology, the complexity and precision required for cancer treatment procedures \cite{xing2006overview, zhang2011parameterization, quan2013preliminary, cao2012uncertainty, an2017robust, liu2018impact, shan2020intensity} require a lot of documentation. Accurate records are essential for multidisciplinary coordination, patient safety, and positive outcomes \cite{roberts2016clinical}. As demand for healthcare services increases, efficient tools are needed to alleviate the documentation burden on healthcare professionals, allowing them to dedicate more time to patient care and other essential healthcare tasks. In radiation therapy CT simulation specifically, therapists manually summarize CT simulation orders into concise notes and document them in the ARIA database (Varian Medical Systems, Palo Alto, CA) or other radiation oncology-specific Electrical Medical Record (EMR) system to help execute the CT simulation. This manual process can be prone to inconsistent writing, inefficient, and burdensome for therapists as different therapists have varying writing styles. It also poses interpretation challenges for research teams, who did not write the notes to understand the writing.

LLMs have shown great capabilities to understand and summarize unstructured texts in healthcare \cite{LIU2023100045, liu2023radonc, LIAO2024128576, shi2024mgh, wang2024recent, rydzewski2024comparative}. To address the aforementioned challenges, in this work, we investigated the use of LLMs to automate the summarization of CT simulation orders, with the goal of reducing variation, improving efficiency, and alleviating clinical workloads. Instead of commercially available LLMs, our study used the locally hosted Llama 3.1 405B model (Meta, Menlo Park, CA) \cite{grattafiori2024llama3herdmodels} to automate the summarization of CT simulation orders from patients while preserving patient privacy.

\section{Materials and Methods}
\subsection{Data} 
\begin{figure}[ht]
\centering
\includegraphics[width=0.8\linewidth]{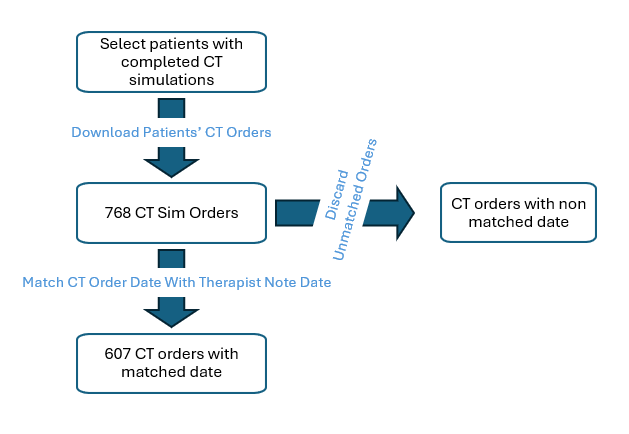}
\caption{\label{fig:processing} Pre-processing the dataset by matching the exam dates. The raw dataset is processed by matching the exam date of the CT simulation orders with the date of the corresponding therapist-wrote notes, retaining only matched CT simulation orders and discarding unmatched CT simulation orders.}
\end{figure}

\begin{figure}[ht]
\centering
\includegraphics[width=\linewidth]{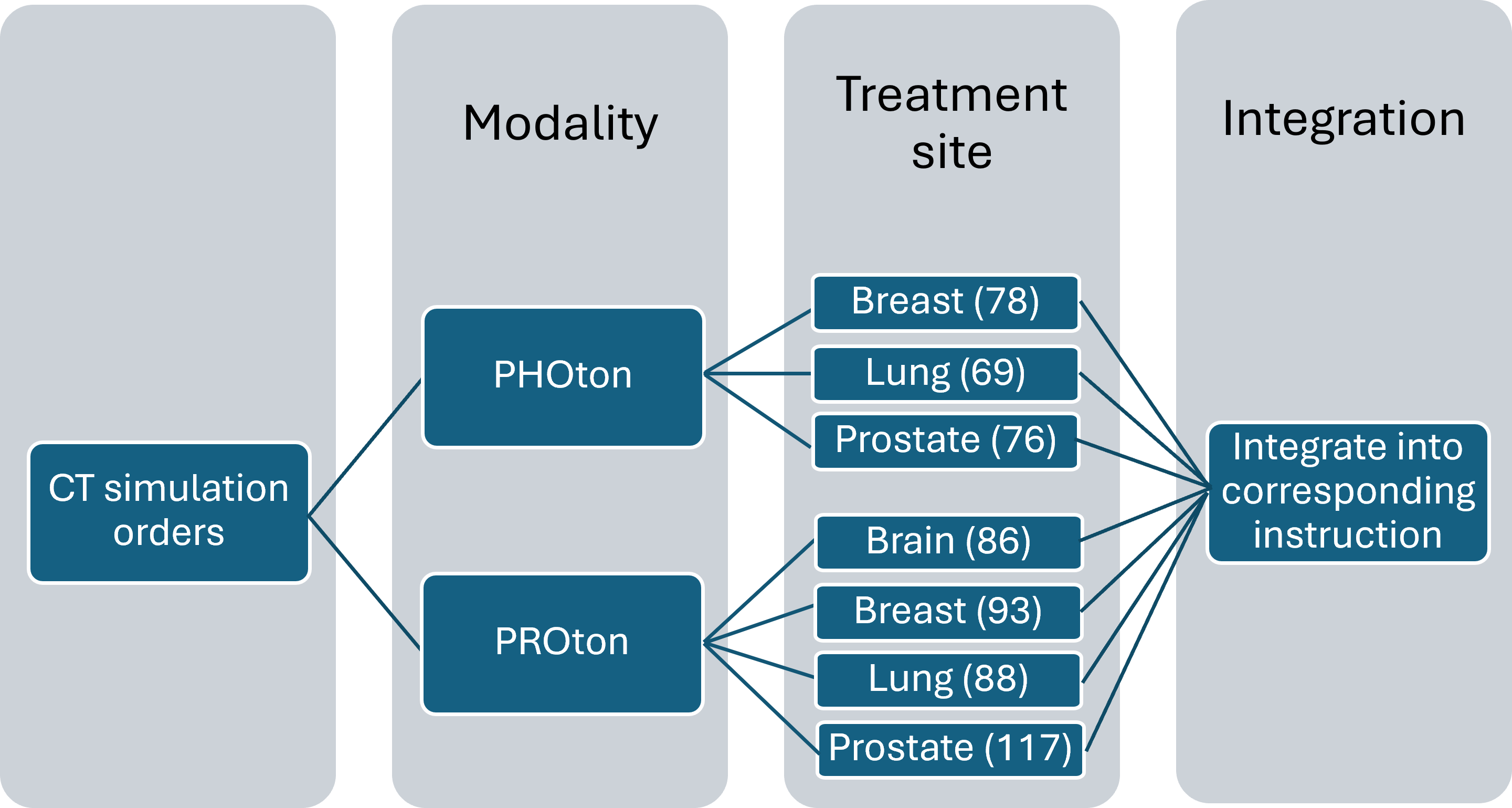}
\caption{\label{fig:selection} Categorization and integration of the dataset. The workflow demonstrates how data is systematically categorized by treatment modalities (such as proton or photon therapies) and disease sites, then data is categorized into 7 groups, ensuring data quality and consistency for analysis.}
\end{figure}

We utilized 768 patient cases whose CT simulations were completed after January 1st, 2019. The CT simulations were retrieved using SQL from the Aria database (ver.15.6) (Varian Medical Systems, Palo Alto, CA) through an in-house patient data look-up tool. 768 patients' CT sim orders and their corresponding therapists' notes were downloaded. Orders were pre-processed using Python to extract the treatment modalities and disease sites. From the 768 CT simulation orders, 7 groups categorized based on treatment modality and disease site were formed: proton (brain, breast, lung, prostate) and photon (breast, lung, prostate). Next, the CT simulation orders within these groups were then matched with the therapists-wrote notes regarding their exam date, further narrowing down the number of samples to 607 CT simulation orders. This final data set was used for further study and the detailed selection process is presented in Fig. \ref{fig:processing}. The number of CT simulation orders across each treatment category is presented in Fig. \ref{fig:selection}.

\subsection{Prompt Engineering} 
To create proper prompts that can summarize CT simulation orders accurately and effectively, general rules and a sample CT order for each category with details were provided by the therapists. To account for different summarization standards across different categories, a customized prompt was developed for each category. Generally, the prompt of a category will include these four parts: (1) set the role of the LLM, (2) provide rules, (3) show examples, and (4) give guidance for the final output. First, the role of the model is defined as a professional medical assistant tasked with assisting the therapists in summarizing CT simulation orders. Second, the prompt provides a structured set of rules to guide the LLM in extracting the required information in a specific sequence and formatting it into standardized medical language. For example, for a CT simulation order specifying the proton modality to treat a lung cancer patient using deep-inspiration breath hold (DIBH), the LLM would first identify the modality and format it as "PROton." Then, it would determine the treatment site and append it with a space, resulting in "PROton Lung." Details like breath management would be added after the treatment site, separated by a comma. In this example, the correct summary would be: "PROton Lung, DIBH." Third, to improve the LLM’s accuracy, a detailed example and its correct summary were included in the prompt. This serves as a reference to ensure consistency and accuracy in the output.  Lastly, the guidelines specify the required output format. Given the tendency of the LLAMA series to explain processes, the rules for this step mandate that the output should concisely be in a JSON format, aligned with the example output. Additionally, the guidelines also emphasize that the LLM should generate the summary without including extraneous phrases or explanations. For example, the output should avoid starting with phrases like "Here is the summary" or describing the process.

\begin{figure}[H]
\centering
\includegraphics[width=\linewidth]{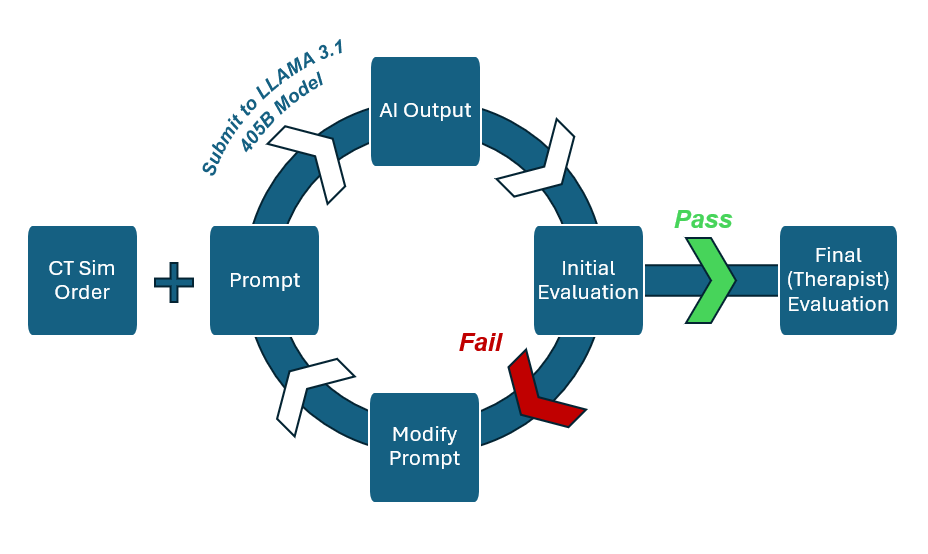}
\caption{\label{fig:promptrefinment}Prompt engineering and evaluation process. The AI output generated from the customized prompt undergoes continuous evaluation until it meets the initial evaluation standards. During this process, the prompt is iteratively refined after each failed evaluation. }
\end{figure}

The prompts were not finished at once. We continuously refine the prompts based on the results generated by the model to increase the accuracy and adapt to the variations of the key information in the CT simulation order, as illustrated in Fig.\ref{fig:promptrefinment}. For example, some CT simulation orders when filled in were not following the regular format. To enhance the model's ability to identify treatment sites, the rules in the prompts were added to include the information listed after "treatment site," "treatment site 1/2," or "anatomical sites." Similar adjustments were made for other required fields to improve the precision and completeness of the CT order summary. Details of the final prompt templates for the 7 categories can be found in the supplementary materials.  

In addition, to reduce the response variation, the temperature was set to 0.1 for LLaMA 3.1 405B. However, different temperature configurations (i.e. 0.7) were explored to double-check the variation of the LLaMA 3.1 405B model's response. For each CT simulation order, the model was queried three times with the same prompt to further check its response consistency.

\subsection{Evaluation}
The AI-generated summaries are evaluated in two steps: first a comparison with ground truth (GT) and then the expert evaluation by therapists. 

\subsubsection{Ground Truth (GT)}
In the first step, the AI-generated summaries are systematically compared against the GT approved by the therapist. The GT was manually created based on three inputs: therapists' notes, CT simulation orders, and therapists' assessments. The irrelevant information of the therapists' notes for each CT order was cleaned up based on the patterns of the notes using Python code and manually. For example, for the original therapist notes - 'Proton left breast, mepitel, teach 9am, mk Dosi: [Dosimetrist name] [initial of the therapist]', the cleaned version would be 'Proton left breast, mepitel', from which the information of MRI and RN appointment, and the notes for therapists themselves were discarded. During manual creation of the GT, the cleaned up therapist notes and the corresponding CT simulation orders were referred simultaneously to ensure consistency and completeness in the data. After that, the manually generated GT was reviewed by therapists to confirm its clinical relevance and accuracy. The finalized GT serve as the benchmark for the initial evaluation of AI-generated summaries, providing a reliable standard for comparison.

\subsubsection{Evaluation using GT}
This step evaluates whether the information in the AI-generated summaries aligns with the one in the GT, including modality, treatment sites, laterality, imaging techniques, mobilization devices, and other critical details. The accuracy threshold is set at 90\%, meaning at least 90\% of the AI-generated summaries must match the GT in terms of completeness and correctness. This step ensures that the AI outputs adhere closely to the intended structure and content. Once the AI-generated summaries meet the 90\% accuracy threshold in comparison with the GT, both the summaries and the GT are sent to an experienced therapist for the final evaluation. 

\subsubsection{Evaluation by therapist}
An experienced therapist reviewed all AI-generated summaries based on his expertise. This step focuses on assessing the clinical relevance, coherence, and overall accuracy of the summaries in the context of real-world healthcare applications. The therapist also evaluates the appropriateness of the information presented in the AI summaries, ensuring alignment with patient-specific circumstances and healthcare system requirements. Accuracies measured in the initial evaluation were reviewed by the therapist during this step.




\section{Results} 

The accuracy and consistency of the LLAMA 3.1 405B model generated results are reported. Accuracy represents the model's performance in adhering to the specified prompt rules, while consistency evaluates its reliability and stability across repeated evaluations under identical temperature settings.

\begin{figure}[H]
\centering
\includegraphics[width=0.8\linewidth]{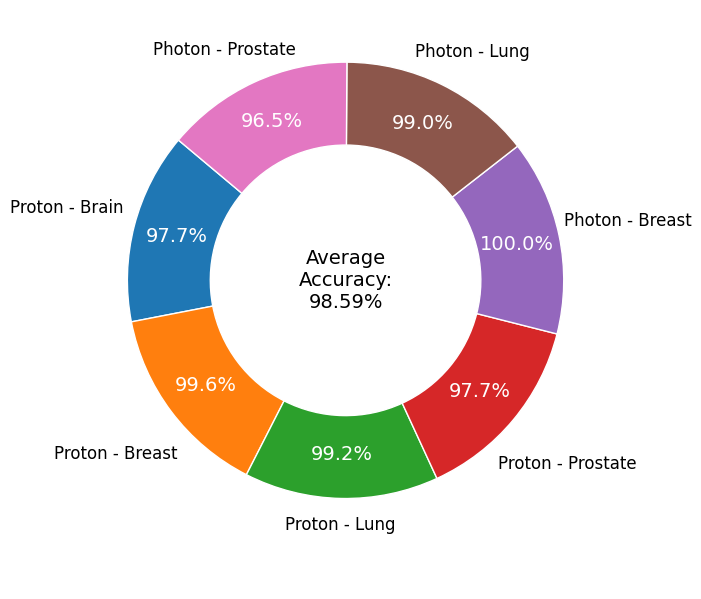}
\caption{\label{fig:results_accuracy}The accuracy of the AI generated summaries across 7 treatment categories. Each color in the circular figure represents a specific category, with the corresponding accuracy of the AI-generated summaries for that category shown in the same color.}
\end{figure}

As shown in Fig. \ref{fig:results_accuracy}, in the final evaluation by the therapist, the generated summaries had an average accuracy of 98.59\% over all seven categories, with photon-breast CT orders having the highest recorded accuracy (100\%). The lowest accuracies were observed for photon-prostate (96.49\%) and proton-brain (97.67\%) CT simulation orders. 

Moreover, consistently high accuracies were observed for the LLAMA 3.1 405B model across different temperature settings. The temperature configuration of 0.1 yields the highest accuracy in summarizing CT simulation orders. Among three trails of the same prompt, the AI-generated summaries also showed great consistency in accuracy (about 98\%). Details of the accuracies obtained at different temperatures, along with the consistency results from three trials, are provided in the supplementary material.  
 
We also tested the performance of the LLaMA 3.1 70B model on this task for a second check. The generated summaries by this model reached an accuracy of 90\% at a temperature configuration of 0.1, lower than the 405B model. Similarly, details of the LLaMA 3.1 70B results on this task are included in the supplementary material.


\section{Discussion}
\subsection{Patient data processing}
In this study, 607 CT simulation orders were included in the final analysis. This number differs from the initially selected orders retrieved from the internal data system due to the MRN number, treatment site, and examination date selection criteria. The CT order selection process is essential for eliminating duplicates, rescheduled CT orders, and orders associated with undesired treatment sites. Multiple treatment sites for the same patient, rescheduling of CT simulations, or other unseen changes in patients' CT simulations can result in multiple CT simulation orders. By matching the patient MRN number, treatment site, and the actual exam date, it verifies that both physician simulation orders and the downloaded therapist notes pertain to the same patient and the same simulation process, and facilitates the grouping of CT simulation orders based on treatment modality and treatment sites.



\subsection{Review of the LLM generated summaries}
While reviewing the summaries generated by the LLaMA 3.1 405B model, subtle differences were observed for a few cases, such as missing the treatment site mentioned in the comment area or inaccuracies in key information. These discrepancies are primarily attributed to ambiguities in the CT orders. For example, for PROton-brain patients' CT orders, instructions regarding bolus can be confusing to the AI model. Normally, the CT order may present the bolus information under the bolus section. However, this information can appear inconsistently, such as "Bolus → Yes; Bolus → No" or "Bolus → Yes; comment → without bolus," or it may be detailed only in the doctor's note section. In this case, the therapists were instructed to do the CT simulation with bolus and repeat the process without bolus. However, additional knowledge in CT simulation is required for the AI model to accurately interpret the information and generate the appropriate summary. This variability in CT orders introduces challenges in the bolus-related sections of the summaries, affecting the model's accuracy and consistency. 

\subsection{Complexity of the clinical data}
Although the overall accuracy of the LLAMA 3.1 405b model exceeds 98\%, categories of PHOton-Breast, PHOton/PROton-Prostate and Proton-Brain exhibit lower accuracies when compared to other categories. This accuracy variation among categories reflects the complexity and variability of the clinical data within each category. For example, within the same category of the CT simulation orders, certain critical details may vary from one to another. When bolus helmet information is present in the PROton-Brain CT orders, it may refer to the mask for the CT simulation process, or the actual bolus helmet used during the radiation therapy. Thus, the actual use of a bolus helmet will need to refer to the patient's clinical history. 

Moreover, the amount of required information in the summarized notes differs across treatment modalities. For example, the PHOton Breast category requires six key pieces of information: Modality, Treatment Sites, Laterality, IV Contrast, Motion Management, and Implanted Device. In contrast, the PHOton Prostate category demands ten pieces of information: Modality, Treatment Sites, Special Instructions, MRI in Treatment Position, Bladder Options, IV Contrast, Treatment Techniques, Chemo Coordination, Motion Management, and Implanted Medical Device. These variations in complexity and specificity of required information could pose a challenge to the LLMs and result in different accuracies across categories, as shown in Fig. \ref{fig:results_accuracy}.

\section{Conclusion}
The results of this study demonstrate the high accuracy and versatility of using an LLM in generating CT order summaries. Our findings showed that LLMs can be potentially integrated into the CT simulation workflow, enhancing consistency, improving efficiency, and reducing the workload associated with the CT simulation order summary task.



\printbibliography

\end{document}